\documentclass[twocolumn,showpacs,%
nofootinbib,aps,superscriptaddress,%
eqsecnum,prd,notitlepage,showkeys,10pt]{revtex4-1}
\usepackage{amsmath,amssymb,amsfonts}
\usepackage{graphicx}
\usepackage{textcomp}
\usepackage{xcolor}

\usepackage[noend]{algpseudocode}
\usepackage{algorithm}
\usepackage{comment}
\usepackage[hidelinks]{hyperref}
\makeatletter 
\newcommand{\linebreakand}{%
\end{@IEEEauthorhalign}
\hfill\mbox{}\par
\mbox{}\hfill\begin{@IEEEauthorhalign}
}
\makeatother 

\def\BibTeX{{\rm B\kern-.05em{\sc i\kern-.025em b}\kern-.08em
		T\kern-.1667em\lower.7ex\hbox{E}\kern-.125emX}}
\begin{document}
	
	\title{
		Unambiguous Interpretation of IEC 60848 GRAFCET based on a Literature Review
	}

	\author{Robin Mroß}
	\affiliation{
		\textit{Lehrstuhl Informatik 11} \\
		\textit{RWTH Aachen University}\\
		52074 Aachen, Germany \\
		\{mross, voelker, kowalewski\}@embedded.rwth-aachen.de
	}
	\author{Aron Schnakenbeck}
\affiliation{
	\textit{Institut für Automatisierungstechnik} \\
	\textit{Helmut-Schmidt-Universität}\\
	22043 Hamburg, Germany \\
	\{aron.schnakenbeck, alexander.fay\}@hsu-hh.de
}
	\author{Marcus Völker}
	\affiliation{
		\textit{Lehrstuhl Informatik 11} \\
		\textit{RWTH Aachen University}\\
		52074 Aachen, Germany \\
		\{mross, voelker, kowalewski\}@embedded.rwth-aachen.de
	}

	\author{Alexander Fay}
	\affiliation{
		\textit{Institut für Automatisierungstechnik} \\
		\textit{Helmut-Schmidt-Universität}\\
		22043 Hamburg, Germany \\
		\{aron.schnakenbeck, alexander.fay\}@hsu-hh.de
	}
	\author{Stefan Kowalewski}
\affiliation{
	\textit{Lehrstuhl Informatik 11} \\
	\textit{RWTH Aachen University}\\
	52074 Aachen, Germany \\
	\{mross, voelker, kowalewski\}@embedded.rwth-aachen.de
}
	
	\begin{abstract}
	IEC 60848 GRAFCET is a standardized, graphical specification language for control functions. Because of the semiformal nature of IEC 60848, the details of specifications created with GRAFCET can be interpreted in different ways, possibly leading to faulty implementations. 
	These ambiguities have been partially addressed in existing literature, but solved in different manners. Based on a literature review, this work aims at providing an overview of existing interpretations and, based on that, proposes a comprehensive interpretation algorithm for IEC 60848, which takes all relevant ambiguities from the literature review into account.
	\end{abstract}

	\maketitle
	%
	
	\section{Introduction}

GRAFCET \cite{iec60848} is a graphical means,
used in education, research, and industry
, to describe, document, and design control behavior in the industrial automation domain. 
GRAFCET evolved from Petri nets in the 70s and became an international standard in 1987 \cite{David.95}. 
Although GRAFCET adapts concepts of Petri nets -~like transitions and steps, connected alternately by arcs~- it provides a considerable number of additional modeling mechanisms like hierarchical structuring of the specification which allow for compact modeling of complex systems \cite{Mross.22}.

The unambiguous interpretation of its syntax and semantics is important for:
\begin{itemize}
    \item its usage as communication means between, e.g., designers and users of automation systems to prevent misconceptions, 
    \item the implementation of tools that allow for a model-driven development to ensure the same behavior of the specification and its implementation, as well as
    \item the verification of the control behavior since the behavior can only be verified if it is unambiguously defined. 
\end{itemize}
However, because of the semiformal nature of the standard, an unambiguous definition of GRAFCET is not given and different authors have interpreted its ambiguities in different ways. In particular, works describing translations from GRAFCET to formal models must make concrete choices on several of the ambiguities discussed in this work during the transformation.
To the best of our knowledge, there exists no documented work with a focus on interpretation of IEC 60848 GRAFCET that is in accordance with the latest version of the standard from 2013.

For other description means from the field of industry automation, there is existing work in order to specify an unambiguous interpretation while trying to include a broader consensus:
von der Beeck \cite{Beeck.94} defined 19 syntactic and semantic problems to compare different Statechart variants that have historically evolved trying to refine its semantics.
Bauer et al. \cite{Bauer.04} presented an interpretation of IEC \mbox{61131-3} Sequential Function Charts (SFC) after comparing the implementation of SFC programming tools of  different Programmable Logic Controller (PLC) vendors.
Wagner et al. \cite{Wagner.08} stated in their interpretation of IEC 61499, that different perspectives and backgrounds of the respective authors like analyzability, performance or usability can lead to different interpretations.

The approach chosen in this work is the identification of ambiguities and its interpretation based on a literature review. The goal of this paper is to propose an interpretation of IEC 60848 GRAFCET in accordance with the latest version of the standard \cite{iec60848}, taking the state of the art into account to resolve its ambiguities.

For the remainder of the paper, we start by pointing out the methodical approach of the literature review we used to identify ambiguities and interpretations of GRAFCET in Sec. \ref{sec:method}. 
The findings are clustered and presented in Sec. \ref{sec:stateOfTheArt}. 
In Sec. \ref{sec:discussion}, we have aimed towards defining a consensus of the observed interpretation by proposing an interpretation algorithm of GRAFCET. We end with a conclusion in Sec. \ref{sec:conclusion}.

Note that due to the limited space, in this publication no descriptions are given of the elements in IEC 60848 and ambiguities are illustrated by small examples, though they also apply to larger GRAFCET instances. For an overview, we refer to \cite{David.95} and \cite{provost.11}.
In the following, the term Grafcet refers to an instance of GRAFCET.
	\section{Methodical approach}
\label{sec:method}
\begin{figure}
	\centering
  \includegraphics[width=0.30\textwidth]{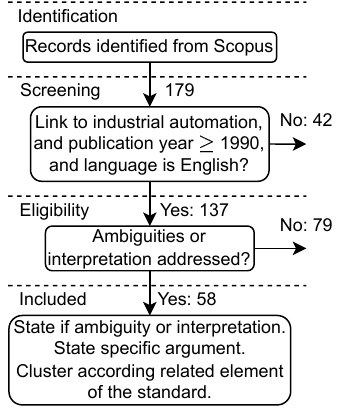}
	\caption{Methodical approach of the literature review.}
	\label{fig:prisma}
\end{figure}

For the interpretation of IEC 60848 we conducted a literature review. 
The literature review consisted of four phases, as shown in Fig. \ref{fig:prisma}. 
We identified 179 titles using the database Scopus\footnote{\url{https://www.scopus.com}} and the search terms \textit{interpret*, synta*, semantic*, analy*, formal*, ambiguit*} or \textit{conflict*} in combination with \textit{grafcet} or \textit{60848}, where one of the latter must appear in the title, abstract, or keywords. The titles were screened, examining the title, abstract and keywords and were excluded if the publication year is before 1990, the language is not English, or the domain is not related to industrial automation. 
The titles were fully read in the third phase and excluded if they address Petri nets or Sequential Function Chart instead of GRAFCET or no ambiguities or interpretations of IEC 60848 are addressed.

Since the goal of this paper is the interpretation of the IEC 60848 based on the version from 2013 \cite{iec60848}, interpretations that are contradicting to the standard from 2013 were excluded from the review as well.  
An example would be the distinction between an interpretation of GRAFCET that searches for stability and one without, made in works during the 1990s, while the standard from 2013 clearly describes transient evolutions which correspond to an interpretation with search for stability.  
Note that not all identified titles are quoted individually when their statements coincided.
For example, a lot of authors introduce GRAFCET with stating consistently that an evolution requires no time, which we considered some kind of interpretation. 
	\section{Results of the literature review and discussion of ambiguities}
\label{sec:stateOfTheArt}
The ambiguities identified in the literature review were clustered and discussed according to the topics presented in the following sections, Sec. \ref{sec:evolutionRules} to \ref{sec:remarks}.

\subsection{Evolution rules and actions}
\label{sec:evolutionRules}
The standard \cite{iec60848} defines five evolution rules that describe how a Grafcet evolves from one situation to the next in a textual way. 
The first rule deals with the initial situation of the Grafcet, which is identified by the set of initial steps. 
The rule fails to state if these initial steps are activated or already active at system initialization. Associated actions on step activation depend on this definition. David et al. \cite{David.95} as well as Sogbohossou et al. \cite{Sogbohossou.15} provide an interpretation algorithm that executes actions on activation associated to initial steps during the initialization. 
Regarding the question of the initial situation's stability, the standard states that initial steps could be unstable, 
which is relevant, e.g., for the evaluation of associated continuous actions. 
The interpretation algorithm by David et al. \cite{David.95} represents this behavior.
Sogbohossou et al. \cite{Sogbohossou.15} state that the initial situation can be transient. 
Additionally, it is unclear how the input variables are initialized and whether or not any rising or falling edge of an input variable can be \textit{true} in the initial instant. 
In the interpretation algorithms by David et al. \cite{David.95} and Sogbohossou et al. \cite{Sogbohossou.15} an external event can only occur if a stable situation has been reached, and this is also true for the initial situation. 
Other authors like Bierel et al. \cite{Bierel.97} propose interpretation algorithms that consider input changes directly after the initial situation is set. We do not consider events before or during initialization in accordance with the immediate execution behavior of GRAFCET, as discussed in Sec. \ref{sec:remarks}.


The second rule defines requirements for a transition to clear in an unambiguous way. 

Evolution rule number three states that by clearing a transition, all preceding steps are deactivated, and all succeeding steps are activated at the same time.
This prevents the perhaps intuitive behavior, that actions on step deactivation happen before actions on step activation, in particular in a chain of steps.

\begin{figure}[ht]
	\centering
  \includegraphics[width=0.2\textwidth]{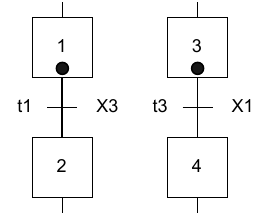}
	\caption{Example of a read-write conflict in the context of evolution rule four.}
	\label{fig:rw-conflict}
\end{figure}
Rule number four defines that transitions that can be cleared simultaneously are cleared at the same time. 
This is reasonable on specification level, but is not feasible for a sequential interpretation algorithm, where transitions are cleared in a sequence. 
This can result in conflicts as shown in Fig. \ref{fig:rw-conflict}. The example shows two transitions that can be cleared simultaneously, since the step variables $X1$ and $X3$ evaluate to \textit{true} ($X1$ and $X3$ are Boolean variables indicating the activation status of step $1$ and $3$, respectively). Depending on the order of execution, this can result in different behavior. When transition $t1$ is cleared, first step 1 is deactivated and $t3$ can not be cleared anymore. A different behavior can be observed if $t3$ is cleared first.
To overcome this issue, authors like Azevedo et al. \cite{Azevedo.99} and Bayó-Puxan et al. \cite{BayPuxan.08} suggest, evaluating the evolutions based on a copy of the system variables. A different approach is to mark transitions that can be cleared before actually clearing them, as presented by Brierel et al. \cite{Bierel.97} and Sogbohossou et al. \cite{Sogbohossou.15}. 
Based on rule three and four, one could argue that also simultaneously executable actions, or combinations of transitions and actions, should execute at the same time, if conditioned actions and transitions conditions evaluate to \textit{true} at the same time. To prevent conflicts here, it is preferable to calculate the assigned values based on a copy of the system variables to simulate a simultaneous execution. The approach of marking the elements to be executed can be problematic when assignments of stored actions depend on each other. 


Rule five aims at resolving a possible ambiguity by executing rule three and four: If an active step is activated and deactivated at the same time, it remains active. This clarifies the resulting activation status of that step, the wording however does not resolve whether the step actually \textit{is} deactivated and activated in the process.   
This is important because, e.g., associated stored actions on step (de)activation either trigger or not, depending on this definition. 
Guéguen et al. \cite{Gueguen.01} specify the definition of (de)activation of steps to resolve this ambiguity: They state that an additional condition for a step being activated is that it is not active. On the other hand, an additional condition for a step being deactivated is that no preceding transition can be cleared. Mallet et al. \cite{Mallet.00} argue similarly. Therefore, when rule five is applied, no activation or deactivation of the associated step is performed.

Note that several authors agree that hierarchical elements have priority over the evolution rules (e.g., \cite{Gueguen.01, Bierel.97}; cf. Sec.~\ref{sec:structuring}).

\subsection{Events}
\label{sec:events}
The standard \cite{iec60848} introduces the concept of input and internal events, which is characterized by the change of at least one value of the respective variables in the form of a rising ($\uparrow$) or falling ($\downarrow$) edge. However, it is not entirely clear from this, if one or a multitude of input events can occur at the same time. This has an impact on, for example, the satisfiability of transition conditions: If at most one input event is present at a time, a condition such as $\mathit{\mathbin{\uparrow}a} \, \mathrm{AND} \, \mathit{\mathbin{\uparrow}b}$ for input variables $a$ and $b$ is never evaluated to \textit{true}. 
David et al. \cite{David.95} argue that due to the immediate processing behavior of GRAFCET, independent events can not occur simultaneously. 
Further, they argue that input events are independent of each other and, therefore, can not occur simultaneously. Several authors like Cassez \cite{Cassez.97} adopt this hypothesis. On the other hand, Guéguen et al. \cite{Gueguen.01} discard this hypothesis and argue that this constraint has no benefit on specification level and causes difficulties on implementation level. For a discussion regarding the implementation of the immediate processing behavior, cf. Sec. \ref{sec:remarks}. 

As an additional ambiguity regarding events, it is pointed out by Guéguen et al. \cite{Gueguen.01} that it remains unclear if an event remains \textit{true} during a transient evolution. 
Are the rising and falling edges caused by an event only available to the first evolution, or until a stable situation is reached if several subsequent evolutions are caused by this input event? 
Guéguen et al. \cite{Gueguen.01} point out that an input event induces a causal chain of atomic evolutions specified by transition clearings according to the evolution rules. A rising or falling edge of an input variable can only be \textit{true} during the first of these atomic evolutions. The standard \cite{iec60848} provides an example of a shift register suggesting the same interpretation, but it is not described in detail.  
One could argue that internal events caused by internal variables should have a similar behavior. 

\begin{figure}[ht]
	\centering
  \includegraphics[width=0.158\textwidth]{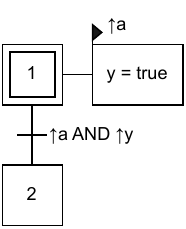}
	\caption{Example Grafcet with a transition condition depending on an input event $\mathbin{\uparrow}a$ and on an internal event $\mathbin{\uparrow}y$, where the former triggers the latter.
    }
	\label{fig:EventIssue}
\end{figure}
\begin{figure}[ht]
	\centering
  \includegraphics[width=0.18\textwidth]{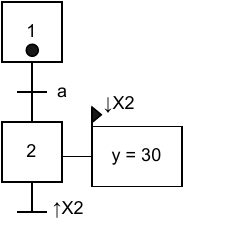}
	\caption{A Grafcet containing an action on event and a transition, both having possibly contradicting conditions depending on rising/falling edges of the step activity variable $X2$. 
	}
	\label{fig:EventIssue2}
\end{figure}

Regarding the definition of events via atomic evolutions mentioned by Guéguen et al. \cite{Gueguen.01} the question arises, how they behave in detail:
If an internal event is triggered by an input event, can both be \textit{true} at the same time?
An example of this is depicted in Fig. \ref{fig:EventIssue}, where the input event $\mathbin{\uparrow}a$ triggers a value change of $y$, resulting in an internal event. 
Here it is ambiguous if $\mathbin{\uparrow}a$ is still present when $\mathbin{\uparrow}y$ arises.
A similar issue arises in the context of step activity variables. 
In Fig. \ref{fig:EventIssue2} step 2 has an action on event associated with it, such that this action depends on the falling edge of the corresponding step activity variable $X2$, it is not clear whether that action will trigger or not if step $2$ is deactivated. 
Likewise, a transition with a condition $\mathbin{\uparrow} X2$  may or may not be cleared, depending on the interpretation. 
We agree that input events are only valid in the first evolution of a transient run, and that internal events happen in a reaction of input events in the context of a casual chain \cite{Gueguen.01} and thereby are \textit{true} in the subsequent evolution. 
Events concerning step activity variables are valid while the respective step is active.

\subsection{Structuring of GRAFCET}
\label{sec:structuring}
The standard defines different notions of plans: connected, partial and global Grafcets.
A connected Grafcet is a plan in which any two elements (steps or transition) are connected by a sequence of arcs. 
The standard states that a Grafcet which is not connected has no technical meaning. However, it is not clear how such a Grafcet would look like. 
Note that a partial Grafcet consists of one or more connected Grafcets 
and does not contain any further partial Grafcets \cite{iec60848}. 
The global Grafcet appears to be unique in a plan and contains partial Grafcets. It remains unclear if a global Grafcet can contain connected Grafcets. 
Regarding this item, the standard states that the situation of a partial Grafcet \textit{G1} can be noted as $G1\{2\}$, provided that only its step $2$ is active. 
This rule talks explicitly about partial Grafcets, so it is not clear which notation to apply for steps that are only part of the global Grafcet, suggesting that the global Grafcet can not contain connected Grafcets. 

The standard \cite{iec60848} 
implies that forcing orders induce hierarchical levels, where a superior partial Grafcet has a higher level than an inferior partial Grafcet. Lesage et al. \cite{Lesage.93} present the resulting hierarchical dependencies as a graph and state that no loop should occur in this graph. Sogbohossou et al. \cite{Sogbohossou.20} formalize these dependencies in terms of a partial order, i.e., the resulting relation has to be transitive, irreflexive and antisymmetric. 
Although Lesage et al. \cite{Lesage.93} do not include enclosings, it is reasonable to ensure a partial order here as well, since cyclic enclosures would result in every corresponding partial Grafcet being active or none of them.

\begin{figure}[ht]
	\centering
  \includegraphics[width=.27\textwidth]{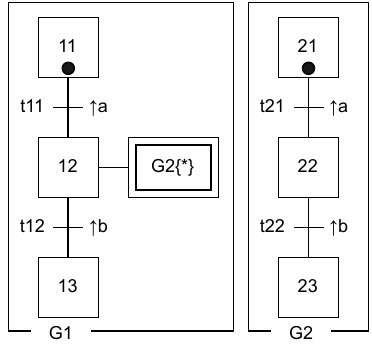}
	\caption{A partial Grafcets $G1$, $G2$, with ambiguous behavior regarding $G1$ forcing $G2$ into its current situation (adapted from \cite{Azevedo.99}).}
	\label{fig:foPriority}
\end{figure}

\begin{algorithm}[H]
	\caption{Possible evaluation of forcing orders I \cite{Azevedo.99}}
	\label {alg:forcingOrder1}
	\begin{algorithmic}[1]
        \ForAll{partial Grafcet (ordered by hierarchical depth)}
            \State Structural evolution
            \State Emit forcing orders
        \EndFor
	\end{algorithmic}
\end{algorithm}
\begin{algorithm}[H]
	\caption{Possible evaluation of forcing orders II \cite{Azevedo.99}}
	\label {alg:forcingOrder2}
	\begin{algorithmic}[1]
        \ForAll{partial Grafcet (ordered by hierarchical depth)}
            \State Structural evolution
        \EndFor
        \ForAll{partial Grafcet (ordered by hierarchical depth)}
            \State Emit forcing orders
        \EndFor
	\end{algorithmic}
\end{algorithm}
Regarding forcing orders, the standard states 
that they have priority over the evolution rules. 
As presented by Azevedo et al. \cite{Azevedo.99} this results in an interpretation algorithm that has to take the hierarchical depth of the partial Grafcet into account (e.g., Lesage et al. \cite{Lesage.93} present an algorithm for the calculation of the hierarchical depth). 
Further, Azevedo et al. \cite{Azevedo.99} point out that two interpretations of this ambiguous statement are possible\label{ref:azevedo}: I) ordered by their hierarchical depth, the new situation of the partial Grafcets is calculated and the forcing orders are immediately applied (cf. Alg. \ref{alg:forcingOrder1}), or II) in a two-step approach, the new situation is first calculated for all partial Grafcets and then the forcing orders are applied for all partial Grafcets again with respect to the hierarchical depth (cf. Alg. \ref{alg:forcingOrder2}). 
The difference regarding the behavior can be illustrated using the example in Fig. \ref{fig:foPriority}, adapted from \cite{Azevedo.99}. 
With the interpretation I) if $\mathbin{\uparrow}a$ appears, $G2$ is immediately forced into the situation $\{21\}$, while with interpretation II) $t21$ can be cleared before $G2$ is forced.
Further, assuming a current situation $\{12, 22\}$ with the first interpretation I) $t22$ can be cleared and with interpretation II) $t22$ can not be cleared, resulting in a situation $\{13, 22\}$. 
Azevedo et al. \cite{Azevedo.99} state that the second approach II), seems to be more comprehensible from the user perspective. However, from our point of view the first approach I) has the advantage of a preemptive behavior which is beneficial to, e.g., implement an emergency stop. Similar arguments can be made for enclosures. We prefer the first approach I), for both hierarchical elements because of the benefits of preemptive behavior.

It is not clear if enclosed partial Grafcets can be controlled by forcing orders, but the standard seems to suggest that they can (cf. Sec. 6.2.2 in \cite{iec60848}). 
However, the designer of the Grafcet should ensure that no conflicts can arise. Possible conflicts are discussed in Sec. \ref{ssec:eclosing}.

\subsection{Macro-steps}
\label{sec:macroSteps}
\begin{figure}[ht]
	\centering
  \includegraphics[width=0.30\textwidth]{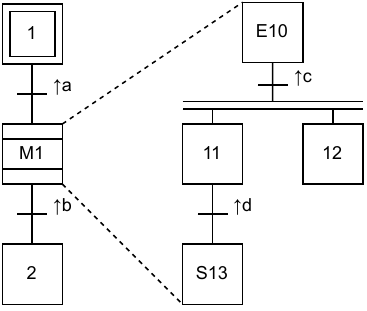}
	\caption{A macro-step and its associated expansion chart, where step $12$ and step $S13$ can be active simultaneously.}
	\label{figM0}
\end{figure}

IEC 60848 introduces macro-steps as an element to structure the Grafcet by means of hiding implementation details of a specification's part in a so-called macro-step expansion chart, e.g., as shown in Fig. \ref{figM0}.
The standard explicitly allows the usage of initial steps within the macro-step expansion chart. Following the rules provided in the standard \cite{iec60848}, the macro-step itself would be active in that case. 
However, an initial macro-step, with a similar symbol to the initial enclosing step, is not proposed by the standard, as pointed out by Sogbohossou et al. \cite{Sogbohossou.20}.

The concept suggests that a macro-step is exited at some point, such that no step within the expansion chart is active until reactivation of the macro-step. However, this is not enforced: It is possible to construct a macro-step expansion chart such that other steps remain active even if the exit step is already activated and the macro-step can thereby be deactivated by its following transition. See for example Fig. \ref{figM0}, where the entry step $E10$ is activated when $M1$ is activated. The sole exit step $S13$ can be active only if step 12 is active as well, and it is unclear if that step remains active when the macro-step $M1$ is exited. In this context, Sogbohossou et al. \cite{Sogbohossou.20} and Årzén et al. \cite{arzen.95} suggest a memory functionality such that these steps regain their previous activity status once the macro-step is reactivated, while being turned off when the macro-step is inactive. Other authors like Wiesmayr et al. \cite{wiesmayr.21} seem to suggest that expansion charts can only be a sequence of steps, circumventing that scenario. In other works, e.g. \cite{Sogbohossou.15}, it is said that macro-steps can simply be replaced with their respective expansion charts, indicating that macro-steps are merely syntactic sugar. The interpretation that macro-steps are elements that can be substituted with their corresponding expansions without semantic loss, as presented in \cite{Sogbohossou.15}, appears reasonable. The behavior regarding steps being active in parallel to the exit step is ambiguous (cf. Fig. \ref{figM0}) and should be avoided by the designer. In regard to the suggested memory functionality \cite{Sogbohossou.20, arzen.95}, there are no references in the standard that support this interpretation.

The standard does not forbid the usage of a source transition in an expansion chart. 
Such a construct could lead to an expansion chart being activated by itself, raising the question of whether this also activates the macro-step. An analog question arises concerning the usage of, e.g., sink transitions, possibly deactivating the entire expansion chart. However, these constructions seem inappropriate for expansion charts and are subsequently not considered.

Another unclarity arises when dealing with nested macro-steps. 
While it is explicitly allowed to have expansion charts that themselves contain macro-steps, it is not clear whether an entry or exit step can be a macro-step. 
The naming convention can not be followed as a step name can not begin with an $M$ and an $E$ or $S$ at the same time, so one can argue that this construction is implicitly forbidden. 
It is further not clear if stored or continuous actions can be associated to macro-steps, as pointed out by Sogbohossou et al.\cite{Sogbohossou.20}. From our point of view, these issues do not result in different behaviors and the choice made here is of little importance.

\subsection{Forcing orders}
\label{sec:forcingOrders}
While it is said for enclosures that an enclosed partial Grafcet can only be associated to one enclosing step, this is not the case for forcing orders. 
A consequence is, that multiple forcing orders can be used to control the same partial Grafcet. 
In scenarios where the mutual exclusion of the steps associated to the orders is ensured, this is no issue. 
However, in complex Grafcets, it might not be easy to determine whether this is guaranteed and the question arises, what happens to the forced partial Grafcet if two of such forcing orders are active at the same time. 
The hierarchical ordering mentioned in Sec. \ref{sec:structuring} can be used to resolve some of these conflicts. When the conflicting forcing orders are emitted from partial Grafcets on different hierarchical levels, the superior one could be prioritized. However, this would lead to a more complex interpretation algorithm, and a possible conflict remains for forcing orders emitted from the same hierarchical level. For this reason, we assume that a Grafcet is designed in a way that no such conflict arises.

\begin{figure}[ht]
	\centering
  \includegraphics[width=0.25\textwidth]{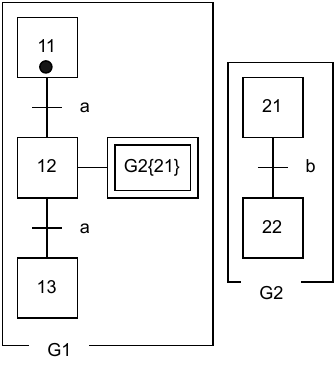}
	\caption{A partial Grafcets $G1$, $G2$, with step $12$ being unstable if $a$ evaluates to \textit{true}.}
	\label{fig:transient_fo}
\end{figure}
Further, the standard \cite{iec60848} states that forcing orders are similar to continuous actions. 
Therefore, the question arises if the behavior regarding transient runs is similar as well: 
Is a forcing order influencing the associated partial Grafcet if the associated step is unstable in the current sequence of evolutions? 
An example is depicted in Fig. \ref{fig:transient_fo} where step $12$ is unstable if $a$ evaluates to \textit{true}.
For example, Wiesmayr et al. \cite{wiesmayr.21} suggest that forcing orders should be executed in a transient evolution. The authors describe how this mechanism can then be used to initialize a partial Grafcet without freezing its situation, allowing it to evolve independently. We prefer this behavior since it can be a useful modeling strategy.


\subsection{Enclosing steps}
\label{ssec:eclosing}
\begin{figure}[ht]
	\centering
  \includegraphics[width=0.20\textwidth]{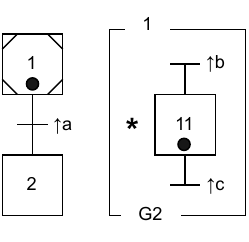}
	\caption{An enclosed partial Grafcet featuring a source and a sink transition.}
	\label{fig5aa}
\end{figure}

The standard states that within each enclosed Grafcet at least one step is active while the enclosing step is as well. 
This statement concerns the question of what can happen to the enclosed Grafcet if the enclosing step is active.
This seemingly implies that this plan can not contain a structure that would make the whole plan inactive, such as sink transitions as shown in Fig. \ref{fig5aa}, where $G2$ is deactivated when $\mathbin{\uparrow}c$ occurs. The standard does not provide syntactical rules to prevent this behavior. 
The reverse imposes interesting questions of similar nature: If an enclosed Grafcet is being activated due to, e.g., a source transition, the partial Grafcet would be active. 
It remains unclear if this is allowed and, if so, if the enclosing step gets activated.
An example is shown in Fig. \ref{fig5aa}, assuming steps $1$ and $11$ are inactive and $\mathbin{\uparrow} b$ occurs. 
 Guéguen et al. \cite{Gueguen.01} state that an enclosed step can not be active if its enclosing step is not.
From our point of view, structures that raise this question should be avoided by the designer.

\begin{figure}[ht]
	\centering
  \includegraphics[width=0.20\textwidth]{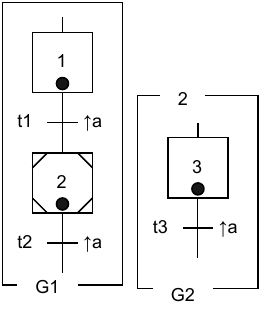}
	\caption{Example regarding ambiguous behavior of enclosings (adapted from \cite{Gueguen.01}).}
	\label{fig:enclosingGueguen}
\end{figure}

Another problem stated by Guéguen et al. \cite{Gueguen.01} is the possible reactivation of the enclosing step, according to evolution rule five, as discussed in Sec. \ref{sec:evolutionRules}: 
It remains unclear if an already active enclosing step will, upon application of the fifth evolution rule, reset the enclosed partial Grafcet to the situation described by activation links (graphically marked with an asterisk) or if it will remain unaffected. Guéguen et al. \cite{Gueguen.01} define additional rules for enclosings specifying that no reactivation of enclosings happens when rule five is applied, similarly to the absent execution of actions on (de)activation as discussed in Sec. \ref{sec:evolutionRules}. One rule defined by Guéguen et al. \cite{Gueguen.01} defines that the enclosed partial Grafcet can not evolve in the same evolution as the succeeding transition of the enclosing step is cleared. 
An example is shown in Fig. \ref{fig:enclosingGueguen}, adapted from \cite{Gueguen.01}. If $\mathbin{\uparrow}a$ arises, evolution rule five is applied to step $2$ and therefore, it stays active. However, according to the additional rule defined by Guéguen et al. \cite{Gueguen.01} $t3$ can not be cleared in the same evolution as $t2$ is cleared.
From our point of view, this restriction seems counterintuitive, since the enclosing step is not deactivated, and the authors do not state why it is necessary. To ensure preemption, the rule could be adjusted so that the enforced partial Grafcet can not evolve if the enclosing step is indeed deactivated, as discussed in Sec. \ref{sec:forcingOrders}.

\begin{figure}[ht]
	\centering
  \includegraphics[width=0.27333\textwidth]{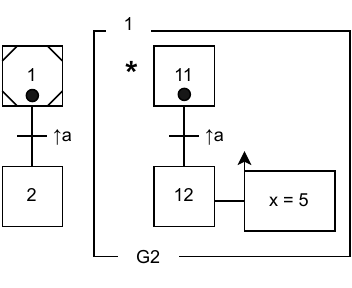}
	\caption{Example Grafcet with ambiguous behavior regarding the preemption of enclosings. 
    }
	\label{fig6aa}
\end{figure}

The standard states that forcing orders have priority with respect to the evolution rules. 
Such a statement is not made for enclosures, which raises a similar question of how to evaluate the transitions properly. Considering the Grafcet in Fig. \ref{fig6aa}, it is not clear whether the action on activation of step~$12$ will execute upon a rising edge of the input variable $a$, or if the enclosed partial Grafcet will be deactivated before. 
The concepts of preemptive behavior, as discussed for forcing orders by, e.g., employing the Alg. \ref{alg:forcingOrder1} or Alg. \ref{alg:forcingOrder2}, could also be applied to enclosures, since  Guéguen et al. \cite{Gueguen.01} state that preemption also applies here, with which we agree.


\subsection{Remarks regarding synchronicity, determinism and algebraic notations}
\label{sec:remarks}
In GRAFCET, evolutions require no time and the number of intermediate step activations and executions of actions required is not bounded. This leads to an issue when such a specification is implemented in a programming language for PLCs. 
The execution mode of PLCs follows the cyclic behavior of reading inputs, executing code and writing outputs. 
All of these parts take longer than zero time, resulting in a discrepancy between the specification and the implementation. For example, two issues are, that the implementation can miss an input signal, or it reacts too late to it. 
Zaytoon et al. \cite{zaytoon.97} discuss different notions of time in that context. 
Other works point out issues regarding implementing a system that has to react to external events, e.g. \cite{LeParc.99}.
They further state that this can be unproblematic if the execution time is considered fast in comparison to occurrences of events in the environment. 

Existing work establishes notions of synchronous, asynchronous and parallel execution models, for example \cite{Zaytoon.00} and \cite{Zaidi.22}, and compares GRAFCET to or provides translation schemes for synchronous or reactive languages \cite{LeParc.93, Panetto.94, Cassez.97, LeParc.99}. 
Provost et al. state in \cite{provost.11} that the evolution rules together with the action definitions of continuous and stored actions on step (de)activations of the standard ensure that determinism of the model is guaranteed. 
The subject of determinism is brought up in other literature as well, in particular in the context of transient runs. 
For example, Carré-Ménétrier et al. \cite{CarreMenetrier.02} assume that orders are only performed when a stable situation is reached, which is not clarified in the standard. 
Cassez \cite{Cassez.97} suggests that outputs are undefined if a stable situation is not present and will not be reached. 
Le Parc \cite{LeParc.99} et al. state that such transient cycles should not be accepted. 
They should instead be detected by, for example, means of verification. 
Several authors propose algebraic notations for GRAFCET and describe aspects of its behaviors by equations, for example in \cite{Panetto.94, provost.11, Perin.13}.

	\section{Interpretation of ambiguities}
\label{sec:discussion}
\begin{figure}[t]
	\centering
  \includegraphics[width=0.33\textwidth]{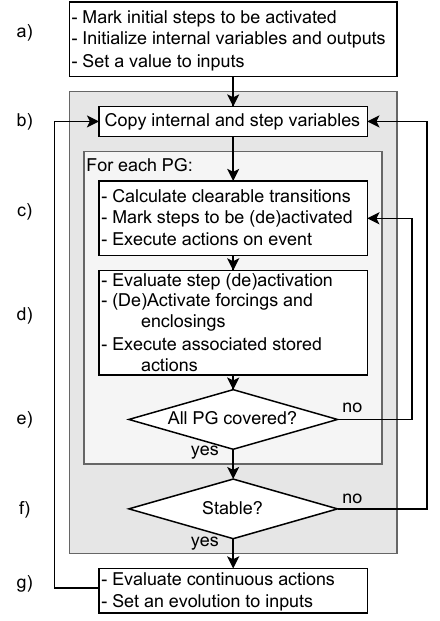}
	\caption{Interpretation algorithm for IEC 60848 GRAFCET (where PG stands for partial Grafcet).}
	\label{fig:algInterpret}
\end{figure}

Several authors present interpretation algorithms for GRAFCET \cite{Sogbohossou.15, Bierel.97, Azevedo.99, BayPuxan.08, Perin.13, Boissier.94, Guignard.13, Julius.19}, the contents of which have been discussed in Section \ref{sec:stateOfTheArt}. 
Based on this discussion, we present an interpretation algorithm for IEC 60848 shown in Fig. \ref{fig:algInterpret}, which addresses the identified ambiguities and solves them in a way respecting the most convincing arguments of the discussion in Sec. \ref{sec:stateOfTheArt}. 


The initial situation is set in step a) in Fig. \ref{fig:algInterpret} and can be unstable, i.e., continuous actions are set after a stable situation is reached even for the initial situation as discussed in Sec. \ref{sec:evolutionRules} and in accordance with the referenced interpretations. No events can occur in the initial situation, since an evolution to inputs happens again after a stable situation is reached in the presented algorithm. This interpretation is in accordance with the initial situation being unstable, since events do not occur during transient runs due to the immediate execution behavior of GRAFCET (cf. Sec. \ref{sec:events} and \ref{sec:remarks}). Finally, because the initial steps are marked as activated, later in step d) the corresponding stored actions on activation will be executed in the first possible evolution. This is in accordance to the findings in Sec. \ref{sec:evolutionRules}.

In the second step b), the system variables are copied to prevent read-write conflicts as suggested in \cite{Azevedo.99, BayPuxan.08} and discussed in Sec. \ref{sec:evolutionRules} and therefore, simulate the behavior that everything happens at the same time in Grafcet as implied by evolution rule four. Note that in the following steps c) and d), all variable values are read from the copied values and written to the original variables. 

Steps c) to e) are executed for every partial Grafcet, ordered by their hierarchical depth, beginning with the  hierarchical highest level as presented in, e.g., \cite{Azevedo.99} (cf. Sec. \ref{sec:structuring}) to ensure preemption. 

In c) a transition can be cleared if all preceding steps are active, the condition evaluates to \textit{true} and the partial Grafcet is not forced according to evolution rule two \cite{iec60848}. 
As discussed in Sec. \ref{ssec:eclosing} by means of the example in Fig. \ref{fig:enclosingGueguen}, the additional rule presented in \cite{Gueguen.01} regarding the clearing of the succeeding transition of an enclosing step is not implemented.

If a transition can be cleared, the steps are marked in order to evaluate evolution rule five in the next step d), and to execute the associated actions, forcing orders and enclosings properly, i.e., to avoid multiple executions or a wrongful activation if the step is already active as discussed in Sec. \ref{sec:evolutionRules}. 
This immediate (de)activation of forcing orders and enclosings results in a preemptive interruption and corresponds to the behavior described in Alg. \ref{alg:forcingOrder1} (cf. Section \ref{ref:azevedo} and the discussion regarding \cite{Azevedo.99}). 
When a forcing is applied, the inferior partial Grafcet has to be somehow marked as forced to prevent the clearing of transitions as demanded by the standard \cite{iec60848}. In order to ensure the priority of forcing and enclosing, the respective steps have to be (de)activated directly instead of using the marker variables. Further, associated stored actions on activation have to be executed. If a step is deactivated by a forcing or enclosing, it remains ambiguous if a possible associated action on deactivation should be executed due to the preemptive manner of forcing and enclosing.
Note that conflicts concerning actions, forcing or enclosing emitted by partial Grafcets with a different hierarchical depth (cf. Sec. \ref{sec:forcingOrders}) are not solved because this would lead to a more complex algorithm. 
Since the hierarchically lower partial Grafcet are executed last in this algorithm, they are  preferred because they would overwrite values. 
We suggest preventing such conflicts entirely using means like verification. 

In step f) an evolution is completed. 
If transitions can still be cleared due to internal events, the evolution is transient as described in \cite{iec60848} and the algorithm proceeds with the next evolution in step b). Regarding the discussion about how long events remain \textit{true} in Sec. \ref{sec:events}, in this interpretation, 
an input event is caused by a value change in step g), or for transient runs an internal event is caused by a value change in steps c) or d) and they remain \textit{true} during the subsequent evolution (indicated by the outer gray box in Fig. \ref{fig:algInterpret}) between steps b) to f). 

In case of a stable situation (step g)), the output variables associated to continuous actions are evaluated and with the next input event the algorithm starts again. 
	\section{Conclusion}
\label{sec:conclusion}
IEC 60848 provides a semiformal language for specifying control functions. Serving as a communication tool, ambiguities in the standard can lead to different interpretations and thereby to divergences in the implementation. This issue is also present when verification approaches tailored to GRAFCET need to be designed, which require some specific interpretation of the standard. The results obtained with these methods can then be subject to an identical understanding of the standard by the users and the engineers of the tool. The standard has been updated to address some unclarities in 2013, providing a more detailed classification of variables. However, literature has pointed out further aspects admitting different interpretations and, in the context of the respective works, the authors have made decisions. In this work we have summarized unclarities in the current version of IEC 60848 GRAFCET and have gathered, where applicable, interpretations in existing literature. Based on this, we have proposed an interpretation algorithm, which again makes decisions on the ambiguous parts of the standard, now based on a review of the body of existing literature and carefully selected interpretations and choices made therein.

	\begin{acknowledgments}
	This research is part of the project Analysis Of GRAFCET Specifications To Detect Design Flaws (project number 445866207) funded by the Deutsche Forschungsgemeinschaft.
	\end{acknowledgments}
	
	\bibliographystyle{IEEEtran.bst}
	\bibliography{bibliography}
	
\end{document}